 \documentclass[final,5p,times,twocolumn]{elsarticle}

\usepackage{graphicx}
\usepackage{epstopdf}
\usepackage{multirow,dcolumn}
\usepackage{rotating}

\usepackage{amssymb}
\journal{Astroparticle Physics}

\begin{document}

\def\nuc#1#2{${}^{#1}$#2}
\def\MJ{{\sc Majorana}}             %Majorana project name
\def\DEM{{\sc Demonstrator}} 
\def\BBz{$0 \nu \beta \beta$}
\newcommand{\gam}{$\gamma$}
\newcommand{\natpb}{$^{\textrm{nat}}$Pb}
\newcommand{\nnprime}{n,n$^\prime \gamma$}

\begin{frontmatter}



\title{Fast-Neutron Activation of Long-Lived Nuclides in Natural Pb}


\author[usd]{V.E.~Guiseppe\corref{cor1}}\ead{vincente.guiseppe@usd.edu}
\author[lanl]{S.R.~Elliott}
\author[lanl,uc]{N.E. Fields}
\author[usd]{D. Hixon}

\address[usd]{University of South Dakota, Vermillion, SD 57069}
\address[lanl]{Los Alamos National Laboratory, Los Alamos, NM 87545}
\address[uc]{University of Chicago, Chicago, IL 60637}

\cortext[cor1]{Corresponding author}

\begin{abstract}
We measured the production of the long-lived nuclides $^{207}$Bi, $^{202}$Pb, and $^{194}$Hg in a sample of natural Pb due to high-energy neutron interactions using a neutron beam at the Los Alamos Neutron Science Center. The  activated sample was counted by a HPGe detector to measure the amount of radioactive nuclides present. These nuclides are critical in understanding potential backgrounds in low background experiments utilizing large amounts of Pb shielding due to cosmogenic neutron interactions in the Pb while residing on the Earth's surface.
By scaling the LANSCE neutron flux to a cosmic neutron flux, we measure the sea level cosmic ray production rates of 8.0 $\pm$ 1.3 atoms/kg/day of $^{194}$Hg, 120 $\pm$ 25 atoms/kg/day $^{202}$Pb, and 0.17 $\pm$ 0.04 atoms/kg/day $^{207}$Bi.

\end{abstract}


\end{frontmatter}

% \linenumbers

\section{Introduction}
\label{sec:Intro}

Neutrinoless double-beta decay (\BBz) plays a key role in understanding the neutrino's absolute mass scale and particle-antiparticle nature~\cite{ell02, ell04, avi08, rod11, bar11, gom12, ell12}. If this nuclear decay process exists, one would observe a mono-energetic line originating from a material containing an isotope subject to this decay mode.
The key to these experiments lies in the ability to reduce intrinsic radioactive
background to unprecedented levels and to adequately shield the detectors from external
 sources of radioactivity. Previous experiments' limiting backgrounds have been trace levels of natural decay chain isotopes within the detector and shielding components. The \gam-ray emissions from these isotopes can deposit energy in the detectors producing a continuum, which may overwhelm the potential \BBz\ signal peak. Great progress has been made in identifying the location and origin of this contamination, and future efforts will substantially reduce this contribution to the background. The background level goal of 1 event/ton-year, however, is an ambitious improvement over the currently best achieved background level~\cite{bau99,gra12,aug12}. If the efforts to reduce the natural decay chain isotopes are successful, previously unimportant components of the background must be understood and eliminated. The contribution from long-lived isotopes produced by cosmic-ray neutrons in detector and shielding materials must be considered. For example, the work of Ref. \cite{ell10} measured cosmic activation of \nuc{76}{Ge}. Several \BBz\ experiments utilize large amounts of low background Pb as a means to passively shield external radiation. If the Pb shielding resides on the Earth's surface prior to deployment at an underground laboratory, cosmic activation occurs. No measurements of cosmic activation in Pb exists or lines identified as such by low background experiments. It is not clear if the lack of prior observations is due to short surface exposures of low background Pb or due to a small activation rate.  Therefore, it is important to understand the production of long-lived activated products in Pb for future \BBz\ decay experiments 
  
We exposed a sample of Pb  to a wide-band neutron beam that resembles the cosmic-ray neutron flux. After exposure we counted the sample in a low-background counting system to observe the \gam\ rays from the decays of  problematic isotopes. From these data we measure the production rate due to fast neutrons in the Pb sample. With knowledge of the neutron-beam and cosmic-neutron energy spectra, we use these data to provide an estimate of the production rate due to exposure of Pb to cosmic rays. We use a cross-section calculation that spans the energy range of interest to validate a comparison of production due to a neutron beam and the most recent cosmic-ray neutron flux measurements of which we are aware. This article describes our determination of  the production rate of these isotopes.

\section{Experiment}

%\subsection{LANSCE WNR GEANIE}

A natural Pb sample was exposed to the neutron beam at the Los Alamos Neutron Science Center (LANSCE) Weapons Neutron Research (WNR) facility from Target 4 Flight Path 60 Right (4FP60R) \cite{lis90}. As the broad-spectrum, pulsed neutron beam strikes the Pb target, the outgoing \gam\ rays are detected by the GErmanium Array for Neutron Induced Excitations (GEANIE) spectrometer \cite{bec97}. The corresponding data from the GEANIE spectrometer and  (n,n'$\gamma$) analysis was presented in a separate publication \cite{gui09}. The GEANIE sample is located a distance of 20.34 m from the natural tungsten spallation target. 

\begin{figure}
\begin{center}
\includegraphics[angle=270,width=0.95\columnwidth]{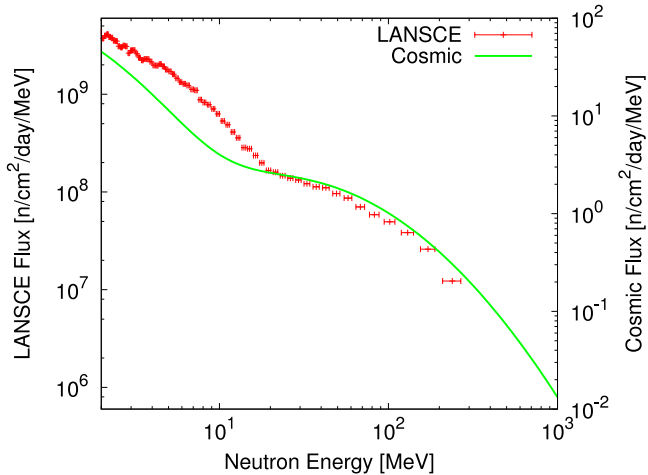}
\caption{The energy spectrum of the neutron beam at 4FP60R for the 2006 exposure periods.  The solid curve represents the sea level cosmic-neutron flux from Ref. \cite{gor04}.}
\label{fig:nFlux}
\end{center}
\end{figure}

The neutron target at the center of GEANIE was five stacked foils of natural Pb (\natpb) angled 20$^{\circ}$ from the normal to the beam direction. Each foil measured nominally 5 cm $\times$ 5 cm in area and 0.475 mm in thickness. The neutron exposure occurred over two periods with 2 foils exposed in the year 2003 and all 5 foils later in year 2006. 
The first 2-foil exposure was performed between Aug. 18 and Aug. 22, 2003 (3.72 days). The full 5-foil exposure occurred between Oct. 13 and Oct. 30, 2006 (16.70 d), Nov. 9 and Nov. 13, 2006 (3.66 d), and on Nov. 22 (0.33 d).  The pulsed neutron beam has the following timing structure. 
Sub-nanosecond-long neutron micropulses occur every 1.8 $\mu$s during a  625 $\mu$s-long macropulse, which occurs at a nominal rate of 40 Hz. The neutron energy is determined by the time of flight from the micropulse start. An in-beam fission chamber measures the neutron flux with $^{238}$U foils \cite{wen93}. 
As seen in Fig.~\ref{fig:nFlux}, the neutron energy spectrum at 4FP60R in 2006 shows good spectral agreement over an energy range between 20-300 MeV with the Gordon \cite{gor04} parameterization of the sea level cosmic-neutron energy spectra.
 If the reader wishes to convolve the neutron spectrum with his/her own cross section model, we give a parameterization of the neutron spectrum impinging upon our sample for convenience. The spectrum can be described as:

\begin{eqnarray}
\label{eq:GEANIENFlux}
\Phi(E) & = & (1.832 \times10^{10}) \times  \\
        &  & e^{(5.166 \ln E - 3.815 \ln^2E+ 0.895 \ln^3E - 0.071 \ln^4E)}  \nonumber 
\end{eqnarray}

\noindent where $\Phi$ is in units of neutrons/MeV and the energy (E) is in MeV.

\begin{figure}
\begin{center}
\includegraphics[width=1\columnwidth]{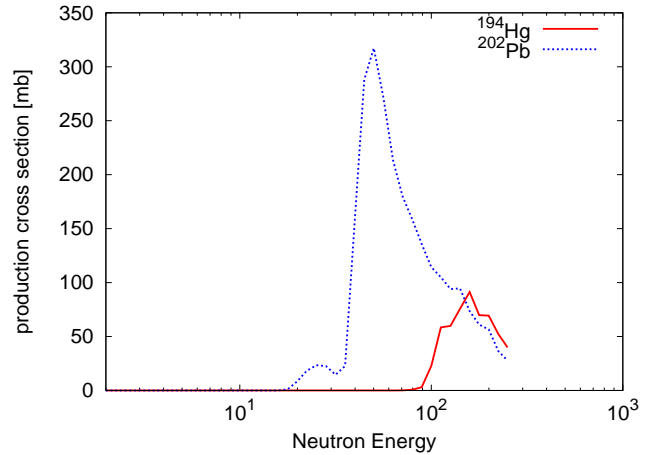}
\caption{The production cross sections for \nuc{202}{Pb} (top curve) and \nuc{194}{Hg} (lower curve) calculated by TALYS for neutrons impinging on natural Pb. The neutron energy range of TALYS is limited to 250 MeV.}
\label{fig:talys}
\end{center}
\end{figure}

The TALYS \cite{kon05} nuclear reaction code was run to to calculate excitation functions for producing nuclides in Pb. Of the nuclides predicted by TALYS, only \nuc{194}{Hg} and \nuc{202}{Pb} are long-lived ($t_{1/2} >$ 1 yr) and support radioisotopes with a decay mode to an excited state so that \gam-ray emission is possible. \nuc{194}{Hg} supports the shorter lived \nuc{194}{Au} ($t_{1/2} =$ 38.0 h) and \nuc{202}{Pb} supports the shorter lived \nuc{202}{Tl} ($t_{1/2} =$ 12.3 d) \cite{nndc}.
The production cross sections for these two nuclides are shown in Fig. \ref{fig:talys}. The production cross section of \nuc{194}{Hg} includes feeding from the short-lived nuclides \nuc{194}{Tl} and \nuc{194}{Pb}. Note that the cross sections are peaked within the energy range of 20-300 MeV, which is the same energy region where there is agreement in the shape of the neutron flux between sea level cosmic neutrons and the 4FP60R neutrons (Fig. \ref{fig:nFlux}).

The experimental setup at GEANIE is optimized for measuring prompt \gam\ rays due to neutron reactions in a target sample. Hence, the neutron irradiation during periods while the data acquisition system is not running does not affect prompt (\nnprime) analysis but could constitute an appreciable neutron fluence for an activation study. Upstream of the WNR target exists a proton current monitor that continuously logs the beam current directed at WNR. This monitor provides a better handle on the total neutron current delivered to the flight path if the fission chamber neutron monitor experiences periods of downtime (e.g. in between data runs, during DAQ malfunctions and debugging, etc.). A subset of the irradiation period while the neutron fission chamber is online can be used to normalize the proton current to delivered neutron current.  

After irradiation, the Pb foils were stored for an extended period and therefore any radioactivity had decayed to an extremely low level 
%($<$150 Bq) 
($<$10 Bq) before counting began. Therefore, this sample was well below any action levels and not subject to any source-handling requirements. The sample was transported to our low-background counting facility underground at the Waste Isolation Pilot Plant (WIPP) near Carlsbad, NM and counted on a HPGe detector. The detector was fabricated in 1985 and placed underground at WIPP in 1998. It is an n-type semi-coax design with a height of 41 mm and a diameter of 51 mm. It is contained within an $\approx$1-mm thick Cu cryostat. The shield during these runs consisted of 5 cm of oxygen-free, high-conductivity Cu and 10 cm of Pb. The detector and its Pb shielding has been underground since 1998.

\begin{figure}
\begin{center}
\includegraphics[width=1\columnwidth]{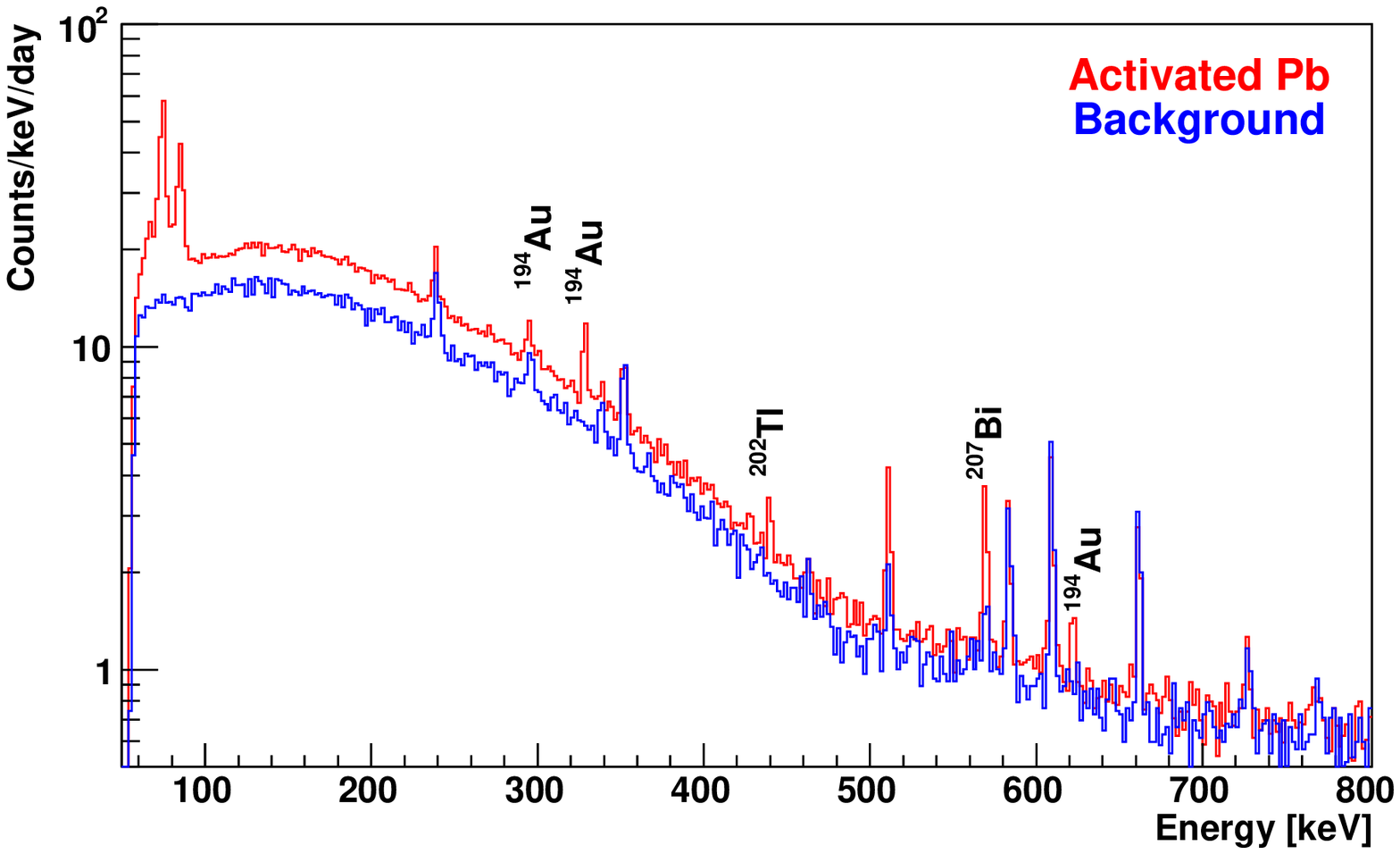}\\
\includegraphics[width=1\columnwidth]{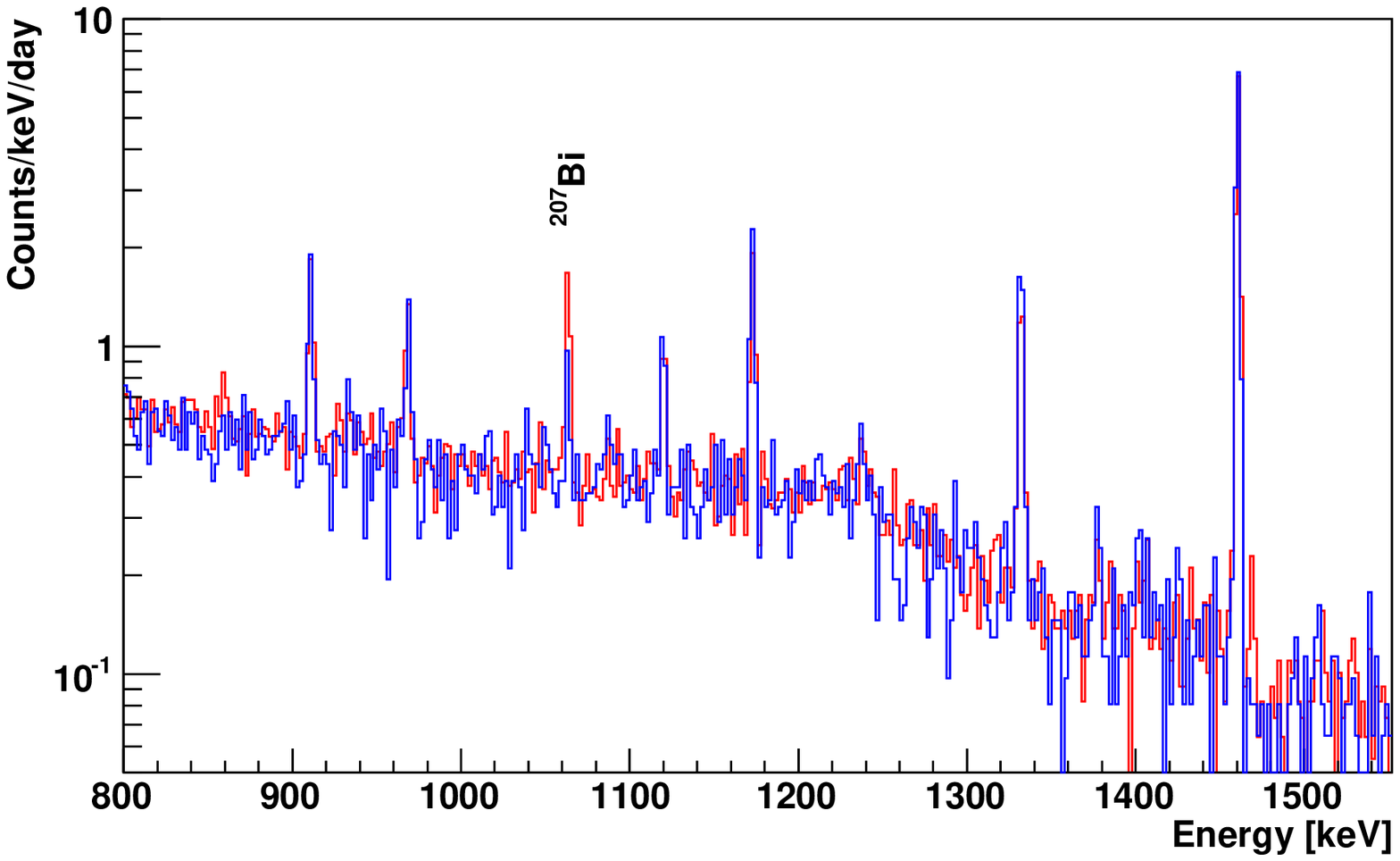}\\
\caption{The energy spectrum of \gam\ rays from the Pb sample (red) as measured by a Ge detector and a background spectrum (blue) taken with no sample present.}
\label{fig:WIPPnSpectrum}
\end{center}
\end{figure}

The sample was counted over a period of 126.6 days between Aug. 4 and Dec. 8, 2010 with a total live time of 109.07 days. The \gam\ ray energy spectrum is shown in Fig.~\ref{fig:WIPPnSpectrum}. Three lines from \nuc{194}{Au}, a line from \nuc{202}{Tl}, and two lines from \nuc{207}{Bi} are observed in the activated spectra in addition to background lines.

\section{Measured Production Rates}
The peaks in Fig.~\ref{fig:WIPPnSpectrum} were fit to determine the measured counts ($C$) and the results are given in Table~\ref{tab:CountResults}. In some of the peaks, the same lines are observed in the background spectra. In the case of \nuc{207}{Bi}, the 569-keV and 1064-keV \gam\ ray lines are also found in the background spectra at a lower rate. The 295-keV \gam-ray line attributed to \nuc{194}{Au} is also found in the background spectra due to naturally occurring \nuc{214}{Pb}. The measured counts in Table~\ref{tab:CountResults} represent the net counts after background subtraction for these three lines. Note that the 328-keV line could be due to decay of naturally occurring \nuc{228}{Ac}. However, the background spectrum shows no evidence of this \nuc{228}{Ac} line and therefore the entire line is attributed to the decay of \nuc{194}{Au}. 

\begin{table*}
%\begin{center}
\caption{A summary of the measured cosmic production rates ($K^C$) and the predicted cosmic production rates ($K^C_{\textrm{\sc talys}}$) for the long-lived isotopes in the Pb sample. Due to multiple \gam\ lines, we report the emphasized rate values given in column 8. If the observed nuclide is short lived, the half-life of the supporting parent (listed in parentheses) is provided. The stated uncertainly on the production rate includes all statistical and systematic uncertainties combined in quadrature. 
%The uncertainty quoted for $N_0$ is a combination of the counting statistics of the measured peaks and the systematic uncertainty in the counting efficiency. 
%Systematic uncertainties are summarized in Table~\ref{tab:CosmicRateUnc} and discussed in the text.
}
\label{tab:CountResults}
%\begin{tabular}{0.99\textwidth}{@{\extracolsep{\fill}} c c c c c c c c c}
\begin{tabular}{c c c c c c c c c}
\hline \hline
Isotope					& $\tau_{1/2}$ 				&  \gam\ ray 		& $C$				& $\epsilon_{\gamma}$	& $\epsilon_{\gamma}$	&  $\epsilon_c$			& $K^C$			& $K^C_{\textrm{\sc talys}}$ 	\\
						& [days]					&  [keV]			& 					& 2006				& 2003					&  					& [/kg/day]			& [/kg/day]				\\
\hline
\nuc{194}{Au} (\nuc{194}{Hg}) 	&	1.90$\times10^5$		&	293.5		& 105$\pm$114		& 0.0041$\pm$	0.0002			& 0.0042$\pm$0.0015	& 3.95$\times10^{-4}$	& 7.0$\pm$7.7		&	\\	
\nuc{194}{Au} (\nuc{194}{Hg}) 	&	1.90$\times10^5$		&	328.5		& 842$\pm$52			& 0.0287$\pm$	0.0014			& 0.0294$\pm$0.0105	& 3.95$\times10^{-4}$	& {\bf8.0$\pm$1.3}	& 16	\\
\nuc{194}{Au} (\nuc{194}{Hg}) 	&	1.90$\times10^5$		&	621.7		& 101$\pm$18			& 0.0009$\pm$0.00005				& 0.0010$\pm$0.0003	& 3.95$\times10^{-4}$	& 29$\pm$7		&	\\
\nuc{202}{Tl} (\nuc{202}{Pb}) 	&	1.93$\times10^7$		&	439.5		& 188$\pm$28			& 0.0444$\pm$	0.0022			& 0.0453$\pm$0.0108	& 3.90$\times10^{-6}$	& {\bf120$\pm$25}	& 77	\\	
\nuc{207}{Bi}			 	&	1.17$\times10^4$		&	567.7		& 316$\pm$37			& 0.0341$\pm$	0.0017			& 0.0346$\pm$0.0055	& 5.94$\times10^{-3}$	& 0.17$\pm$0.03	&	\\	
\nuc{207}{Bi}		 		&	1.17$\times10^4$		&	1063.7		& 146$\pm$26			& 0.0153$\pm$	0.0008			& 0.0153$\pm$0.0011	& 5.94$\times10^{-3}$	& {\bf0.17$\pm$0.04}&	\\
\hline \hline
\end{tabular}
%\end{center}
\end{table*}

The counting efficiency was determined by Monte Carlo simulations using the GEANT4-based MaGe framework \cite{bos11}. Simulations of the detector's counting efficiency is benchmarked against measurements of calibrated radioactive standards (\nuc{57,60}{Co}, \nuc{54}{Mn}, \nuc{22}{Na} and \nuc{137}{Cs}). 
%Agreement between data and Monto Carlo is within 5\%, which constitutes the main systematic uncertainty in the counting efficiency. 
We use a random pulser to verify the event-rate dependence of the data-acquisition system dead time.

In addition to the \gam-ray detection efficiency ($\epsilon_{\gamma}$), an efficiency factor must also be included to take into account the decay of the isotope since the end of exposure and during the counting period. This latter efficiency factor ($\epsilon_c$) depends on the half-life of the isotope of interest and is also given in Table~\ref{tab:CountResults}.

With these efficiencies, it is straight-forward to calculate the number of atoms of each isotope that were present at the end of neutron exposure (November 22, 2006) and the production rate. 

The measured number of counts ($C$) is related to the number of atoms ($N_i$) on the reference date at the end of exposure by:

\begin{equation}
\label{eq:counting}
C = \epsilon_c \sum_i N_i \epsilon_{\gamma_i},
\end{equation}

\noindent where the sum is over the four irradiation periods. Since the first irradiation period only included two of the Pb foils, atoms produced from that run have a different \gam-ray detection efficiency than those produced in the latter three runs with all five Pb foils. The decay of the isotope before and during counting is corrected by

\begin{equation}
\label{eq:epsilonC}
 \epsilon_c = \sum_j \left( e^{-\lambda T^{start}_j} - e^{-\lambda T^{end}_j} \right),
 \end{equation}

\noindent where $T^{end}_j$ ($T^{start}_j$) is the number of days since the end of exposure that the counting stopped (started) for each of the $j$ data runs, and $\lambda$ is the decay constant of the isotope in question.

The number of atoms can also be expressed in terms of the reaction rates during an exposure period:
\begin{equation}
\label{eq:exposure}
N_i = \frac{K^L M_i}{\lambda} \left(1 - e^{-\lambda T^{irrad}_i} \right) e^{-\lambda T^{decay}_i}
\end{equation} 
\noindent where we have corrected for the decay during the exposure and the decay after the exposure until the reference date.
 In Eqn.~\ref{eq:exposure},  $T^{irrad}_i$ is the duration of irradiation of the $i^{th}$ exposure,  and $T^{decay}_i$ is the time between the end of exposure $i$ and the reference date. $K^L_i$ is the production rate (atoms/kg/day) during the LANSCE exposure $i$, and $M_i$ is the mass of the sample within the beam spot.
 
The production rate of the identified long-lived nuclides in Pb occurs in an energy range where the shapes of the LANSCE and cosmic neutron fluxes are similar. Hence, a simple scaling of the LANSCE production rate is sufficient to estimate the cosmic production rate, given by
  \begin{eqnarray}
 K^C  & = & K^L_i  \frac{f^C}{f^L_i}\\
       	& = & K^L_i  \frac{f^C}{S_i I_i / A_i / T^{irrad}_i}
 \end{eqnarray}
  \noindent where $K^C$ is the cosmic production rate (atoms/kg/day), $f^C$ ($f^L$) is the energy-integrated neutron flux for cosmic (LANSCE) neutrons over the mutual energy range 20-300 MeV, $S_i$ is the normalization between proton current and neutron current, and $A_i$ is the cross-sectional area of the Pb target in the neutron beam. The sea-level cosmic neutron flux of Ref. \cite{gor04} is assumed where $f^C$ = 228 n/cm$^2$/day.

Eqn.~\ref{eq:exposure} can be written as
\begin{equation}
\label{eq:exposure2}
	C = \frac{K^C \epsilon_c}{f^C \lambda} \sum_i \frac{n_i S_i I_i}{T_{irrad}^i } \epsilon_{\gamma_i} \left(1 - e^{-\lambda T^{irrad}_i} \right) e^{-\lambda T^{decay}_i}
\end{equation} 
 \noindent where the $M_i/A_i$ term  is replaced by the areal density ($n_i$) of the Pb target parallel to the beam direction. 
The beam spot covers most, but not all, of the sample geometry.

Table \ref{tab:CountResults} lists the isotopes found and the number of measured counts after irradiations. Eqn. \ref{eq:exposure2} allows a direct determination of the cosmic production rate $K^C$, also listed in Table \ref{tab:CountResults}.
Of the three \nuc{194}{Au} lines observed, two have low branching ratios and so we only place the most confidence on the 238-keV line. Further, the 622-keV line can occur as a lone \gam\ ray or as a sum of the 293.5-keV and 328.5-keV \gam\ rays. Its excess may be due to an additional systematic in Monte Carlo of the \gam\ ray efficiency. Further, there may be presence of the 622-keV \gam\ ray from \nuc{106}{Rh}, which is progeny of fission product \nuc{106}{Ru} predicted by TALYS to be activated at a rate of 0.05 atoms/kg/day. If we assume the excess of the 622-keV line is due to \nuc{106}{Rh}, we find a cosmic activation rate of 0.18$\pm$0.06 atoms/kg/day for that isotope. 
The two \nuc{207}{Bi} lines are found to have consistent production rates.

Also listed is the cosmic activation rate calculated with the TALYS excitation functions (Fig. \ref{fig:talys} and the sea level cosmic neutrons \cite{gor04}). 
The production of \nuc{207}{Bi} in Pb requires an increase in the atomic number from secondary reactions from incident neutrons. Hence, the production is not predicted by TALYS.

The aerial densities and total neutron yield for the two irradiation periods is listed in Table \ref{tab:ndata}. The low neutron yield and lower Pb thickness in the 2003 run make it a small contribution to the total production of nuclides.

\begin{table}[ht]
\begin{center}
\caption{A summary of the total neutron yield for the two irradiation years.}
\label{tab:ndata}
\begin{tabular*}{0.75\columnwidth}{@{\extracolsep{\fill}} c c c  }
\hline \hline
Year		&	$n$ 			& Neutrons on target (S $\times$ I)	\\
		&	(g/cm$^2$)		& between 20-300 MeV	\\
\hline
2003		&	1.215			& 5.73$\pm5.46\times10^{10}$		\\
2006		&	3.038			& 6.77$\pm0.28\times10^{11}$		\\
\hline \hline
\end{tabular*}
\end{center}
\end{table}

The uncertainty in the production rate is dominated by the counting statistics of weak peaks (Table \ref{tab:CountResults}), the counting efficiency, the exposed neutron yield, and the cosmic ray neutron flux. The \gam-ray efficiency for the full five Pb foils exposed in 2006 is known to 5\%, which is based on source data and Monte Carlo agreement. The placement of the two Pb foils exposed in 2003 within the five foils during \gam\ counting is unknown, though they were adjacent. Therefore, the 2003 \gam-ray efficiency has a larger systematic uncertainty (Table \ref{tab:CountResults}). 
The uncertainty on neutron yield is due to the statistical uncertainty on the proton beam current and the uncertainty on the proton to neutron normalization. The 2006 total neutron yield is known to 4.2\%. For the 2003 neutron yield calculation, the limited fission chamber data available suggests a lower proton to neutron normalization, though no physical explanation can explain the discrepancy with the 2006 normalization. We assume the nominal 2006 normalization for the 2003 neutron yield, but conservatively include a larger uncertainty of 95\% to remain consistent with the limited data. The 2003 exposure represents $3.3\%$ of the total neutron exposure so the large uncertainty on the 2003 neutron yield has a small effect on the net production rate. 

The precision to which the cosmic ray neutron flux is known is 10-15\% and so we split the difference and use 12.5\% \cite{gor04}. The total cosmogenic rate includes contributions from subdominant proton and pion interactions. These only contribute approximately 10\% \cite{bara06} to the total production rate. These charged particles are much less penetrating than neutrons and therefore their impact on any given sample is very geometry dependent. Hence we assume a 50\% uncertainly on this correction for an uncertainty of 5\%. Again these uncertainties are uncorrelated and result in a estimated systematic uncertainty on the cosmic neutron flux of 13.5\%. The start and stop times of counting and the live time of the counting are known to a small percentage and are negligible contributions to the uncertainty. This is similar for the times associated with the irradiation.  The sample had been stored on the Earth's surface for many years prior to exposure to the beam and then counting at WIPP. Any isotopes produced by cosmic ray neutrons would certainly be below saturation after this extended period with a cosmic contribution $\ll3\%$ . The saturation production rates however are expected to be low. Therefore the total count rate is expected to be dominated by the intense neutron beam and we ignore this systematic effect. The values for the half-lives and branching ratios are known to high precision and are a negligible contribution to the total uncertainty. 

% XXXX do we want an uncertainty table?
%\begin{table*}[t]
%\caption{A summary of the uncertainties (in \%) that contribute to the total uncertainty of cosmic production rate. 
%\label{tab:CosmicRateUnc}
%\begin{tabular}{|c|c|c|c|c|c||c||c|c|c||c||}
%\hline  
%Isotope				& Counting 		&	Efficiency &  Source    & Predicted	&Flux			& SubTotal	& Cosmic	 	&Neutron		&Proton        &Total \\
%(Line Energy)		& Statistics		&			   &  Activity  & 4FP60R 	 & Chamber      &           &  Neutron 	& Specral    &Correction    &\\
   %           		&				&			   &              & Fluence    &Live Time      &           &  Flux      &Difference  &    &\\
%\hline\hline
%\nuc{57}{Co}(122)		&2.9			&0.4	        &1.0        &0.9             &0.3            &3.2       		 &12.5  	& 49        &5.0        	&51 \\
%\\hline
%\end{tabular}
%\end{table*} 

\section{Discussion and Conclusion}

We measured the production of \nuc{194}{Hg}, \nuc{202}{Pb}, and \nuc{207}{Bi} in a sample of Pb due to high-energy neutron interactions within a neutron beam with a spectrum similar to that of the cosmic-ray neutron flux at the Earth's surface. These results can be used to predict sea level neutron activation of long-lived nuclides in Pb shielding. Previous low background experiments using Pb shielding have not found lines in their detectors energy spectrum attributed to long-lived activated products in Pb. For example, the HEILDELBERG-MOSCOW (H-M) \cite{kla01} neutrinoless double-beta decays experiment does not present evidence of \nuc{194}{Au} or \nuc{202}{Tl}. The H-M spectra does show the presence of \nuc{207}{Bi}, though its attributed to anthropogenic rather that cosmic production in Pb shielding. The \nuc{207}{Bi} may be partially cosmogenic in origin in the Pb shielding. If cosmogenic, the decay rate of \nuc{207}{Bi} observed here suggests that  \nuc{194}{Au} and \nuc{202}{Tl} would also be observed. However, the activation of \nuc{207}{Bi} here is due to secondary reactions involving protons and the net cosmic production would be greater due to the flux of cosmic protons.

The work of Mei \emph{et al.} \cite{mei08} using a Pb shield surrounding a HPGe detector did observe a line at 438.9 keV, which is suspiciously close the the \nuc{202}{Tl} line at 439.5 keV observed here. A long term spectra obtained by Ref. \cite{bos05} also observed a line at 438.8 keV but attributed it to a \nuc{40}{K} double escape peak. The two also see the 328 keV line from \nuc{228}{Ac}, which could be blending with cosmogenic \nuc{194}{Au}. However, the rate of the 328 keV line is consistent with the nearby \nuc{228}{Ac} 338 keV line when scaled by their branching ratios.

Neutrinoless double-beta decay experiments typically choose target nuclei with a high ($\geq 2$ MeV) double-beta Q value in order to stay above most \gam\ rays from natural sources. The activated products discovered here do pose a potential background contribution due to their high decay Q values and high energy \gam\ rays. While \nuc{194}{Au} has a decay Q-value of 2501 keV with \gam\ rays up to 2413 keV and  \nuc{207}{Bi} has a decay Q-value of 2398 keV with \gam\ rays up to 1770 keV, \nuc{202}{Tl} is less of a concern with a lower Q value of 1363 keV \cite{nndc}. The production rates of these cosmically activated radioisotopes in Pb can be used to estimate background contributions and surface exposure limitations for the next generation neutrinoless double-beta decay experiments.

With its high energy \gam\ rays, \nuc{194}{Au} is likely the most problematic cosmogenic in Pb shielding. The highest intensity \gam-ray line above 2 MeV occurs at 2043.7 keV. This transition is close to the 2039-keV  Q value of double-beta decay target \nuc{76}{Ge}. Using the production rate measured here, cosmogenic \nuc{194}{Au} will saturate at a concentration of 93 $\mu$Bq/kg. With a decay intensity of 3.81\%, the 2044-keV \gam\ ray would be produced at a rate of 3.5 $\mu$Bq/kg at saturation. As stated earlier, the H-M experiment did not see evidence of \nuc{194}{Au}. To estimate their sensitivity to a 2 MeV \gam\ ray, they did observe the 2.6-MeV \gam\ ray from \nuc{208}{Tl} in the \nuc{232}{Th} decay chain and determined the concentration to be 12.3 $\mu$Bq/kg of \nuc{232}{Th} (4 $\mu$Bq/kg of \nuc{208}{Tl}) in the Pb shielding \cite{dor03}. Therefore, they would have sensitivity to \nuc{194}{Au} at saturation indicating the Pb shielding had little exposure to cosmic neutrons at the surface of the Earth prior to use underground. Current and future experiments are finding shielding materials with improved contamination levels. Pb suitable for shielding has been obtained at $< $4 $\mu$Bq/kg of \nuc{232}{Th} ($< $1.3 $\mu$Bq/kg of \nuc{208}{Tl}) \cite{leo08}. Since  \nuc{232}{Th} contamination in Pb is currently unavoidable, it can be used to set the scale of other contaminate goals. In order to keep the decay rate of the 2044-keV \gam\ ray at or below the decay rate of the \nuc{208}{Tl} 2614-keV \gam\ ray, Pb surface exposure should be limited to 350 yr. A limitation of 30 yr of surface exposure is required to stay a factor of 10 less than the \nuc{208}{Tl} decay rate.

\section{Acknowledgments}
 We gratefully acknowledge the support of the U.S. Department of Energy through Award Number DE-SCOO05054, the LANL LDRD Program, and the  DOE/NNSA Stewardship Science Graduate Fellowship Program under grant number DE-FC52-08NA28752 for this work.
 This work benefited from the use of the Los Alamos Neutron Science Center, funded
 by the U.S. Department of Energy under contract DE-AC52-06NA25396. 
 This work also benefited from our underground laboratory at the Waste Isolation Pilot Plant (WIPP), which we operate with support from the Nuclear Physics office of the U.S. Department of Energy under contract number 2011LANLE9BW. Finally, we thank our friends and hosts at the Waste Isolation Pilot Plant (WIPP) for their continuing support of our activities underground at that facility.

%% References


\bibliographystyle{model1-num-names}
\bibliography{/Users/vguiseppe/work/latex/bib/mymj}






% \end{thebibliography}

\end{document}